\documentclass[prb,twocolumn,showpacs]{revtex4}
\usepackage{graphicx} 
\usepackage{amsmath,amsthm,amssymb,mathrsfs}
\usepackage{bm}

\begin{document}

\title{Spin torque oscillator for microwave assisted magnetization reversal}

\author{Tomohiro Taniguchi and Hitoshi Kubota}
 \affiliation{
 National Institute of Advanced Industrial Science and Technology (AIST), Spintronics Research Center, Tsukuba, Ibaraki 305-8568, Japan 
 }

 \begin{abstract}
{
A theoretical study is given for the self-oscillation excited in a spin torque oscillator (STO) consisting of 
an in-plane magnetized free layer and a perpendicularly magnetized pinned layer in the presence of a perpendicular magnetic field. 
This type of STO is a potential candidate for a microwave source of microwave assisted magnetization reversal (MAMR). 
It is, however, found that the self-oscillation applicable to MAMR disappears when the perpendicular field 
is larger than a critical value, which is much smaller than a demagnetization field. 
This result provides a condition that the reversal field of a magnetic recording bit by MAMR in nanopillar structure 
should be smaller than the critical value. 
The analytical formulas of currents determining the critical field are obtained, 
which indicate that a material with a small damping is not preferable to acheive a wide range of the self-oscillation applicable to MAMR, 
although such a material is preferable from the view point of the reduction of the power consumption. 
}
 \end{abstract}

 \maketitle




\section{Introduction}
\label{sec:Introduction}

A spin torque oscillator (STO) \cite{kiselev03,slavin09,bertotti09text} has attracted much attention from the perspectives of fundamental magnetism and nonlinear science, as well as applied physics. 
The advantage and uniqueness of STO lie in its sub-micron or even smaller component size, wide frequency tunability, high compatibility with current technology, 
and the possibility to be applied to new technologies such as neuromorphic architectures \cite{locatelli14,grollier16,kudo17,torrejon17}. 
In particular, an STO consisting of an in-plane magnetized free layer and a perpendicularly magnetized pinned layer 
\cite{kent04,lee05,houssameddine07,firastrau07,ebels08,silva10,igarashi10,suto12,lacoste13,kudo14,kudo15,bosu16,hiramatsu16,taniguchi16,taniguchi16APEX,bosu17} 
is a potential candidate as a microwave generator for magnetic recording head based on the technology using microwave assisted magnetization reversal (MAMR) 
\cite{bertotti01,thirion03,denisov06,sun06,zhu08,okamoto08,bertotti09,okamoto10,barros11,okamoto12,okamoto13,tanaka13,barros13,cai13,okamoto14,taniguchi14,suto15,taniguchi15APEX1,taniguchi15APEX2,suto16,taniguchi16PRB,suto17}. 
MAMR is a newly proposed technique of magnetization reversal in a nanostructured ferromagnet, 
where the microwaves excite an oscillation of the magnetization in a recording bit, resulting in a reduction of a direct magnetic field necessary to reverse the magnetization. 
Recently, MAMR was demonstrated experimentally \cite{okamoto12,okamoto13,suto17}, 
where linearly or circularly polarized microwaves were generated from electric currents passing through coplanar wave guides. 
A currently exciting topic in MAMR is to replace the coplanar wave guides with the STO for high-density magnetic recording \cite{zhu08}. 
This type of STO can emit circularly polarized microwaves with clockwise or counterclockwise chirality, depending on the direction of electric currents injected into the STO, 
which is a convenient scheme for designing MAMR architecture \cite{kudo15}. 


Substantial efforts have been made to confirm MAMR in practice, as well as reading bit information from the frequency shift, 
with an STO consisting of an in-plane magnetized free layer and a perpendicularly magnetized pinned layer 
\cite{suto12,hiramatsu16}. 
In designing high density recording media by applying MAMR technology for commercial use, the bit will be aligned parallel to the current.
Currently, however, MAMR is investigated in a nanopillar structure 
having a perpendicularly magnetized ferromagnet fabricated onto the STO. 
This perpendicularly magnetized ferromagnet corresponds to the recording bit. 
An external perpendicular field is applied to the structure, which acts as a reversal field of the bit. 
The stray field from the STO plays the role as a microwave field and assists the reversal. 
However, a magnetization reversal assisted by microwaves generated from this type of STO has not been clearly observed yet in the nanopillar structure. 
It is unclear whether there are certain physical conditions and/or restraints on slecting materials, sample structures, etc., in order to observe MAMR by STO. 
Therefore, it is highly desirable to clarify the oscillation properties of STO which can be applied to MAMR observation.

In this paper, we study the magnetization dynamics excited in this type of STO in the presence of a perpendicular magnetic field 
by solving the Landau-Lifshitz-Gilbert (LLG) equation both numerically and theoretically. 
We show that three phases of the magnetization dynamics, i.e., 
an in-plane magnetized state, out-of-plane self-oscillation state, and perpendicularly magnetized state, appear at zero temperature, 
depending on the current and applied field. 
It is found that the self-oscillation state applicable to MAMR disappears when a perpendicular field is larger than a critical value. 
The result provides a condition that the reversal field of a magnetic recording bit should be smaller than 
the critical value to observe MAMR in nanopillar structures. 
We also derive analytical formulas of currents characterizing the different phases of the magnetization dynamics. 
The analytical formulas indicate that a material with a small damping is not preferable for a free layer of the STO 
to achieve a wide range of the self-oscillation applicable to MAMR, 
although such a material is preferable from the view point of the reduction of the power consumption. 
We also perform numerical simulation of the LLG equation in the presence of thermal fluctuation, 
and confirm that the disappearance of the self-oscillation is found even at finite temperature. 


The paper is organized as follows. 
In Sect. \ref{sec:System description and LLG equation}, 
we show the results of the numerical simulation of the LLG equation at zero temperature. 
In Sect. \ref{sec:Numerical simulation at finite temperature}, 
we confirm that the results obtained at zero temperature are still valid at finite temperature. 
In Sect. \ref{sec:Conditions to observe MAMR}, we discuss the conditions to observe MAMR 
based on the results obtained in the previous sections. 
The conclusions of this work is summarized in Sect. \ref{sec:Conclusion}.



\section{System description and LLG equation}
\label{sec:System description and LLG equation}

In this section, we describe the details of an STO in this study, 
and show both the numerical and analytical solutions of the LLG equation of the STO at zero temperature. 
The relation between the oscillation properties of the STO and MAMR is discussed in Sect. \ref{sec:Relation to MAMR}. 
An important conclusion found in Sects. \ref{sec:Theoretical formulas of characteristic currents} and \ref{sec:Critical field}
is that the out-of-plane self-oscillation state applicable to MAMR disappears when the magnitude of the applied field is larger than a critical value. 




\begin{figure}
\centerline{\includegraphics[width=1.0\columnwidth]{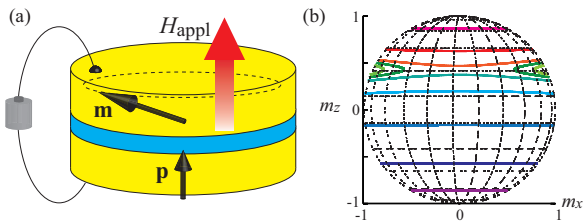}}
\caption{(Color online)
        (a) Schematic view of the system in this study. 
            The current in this figure corresponds to a negative current, where the electrons flow from the pinned to free layer. 
        (b) Schematic view of the constant energy curves of the free layer, where $H_{\rm appl}=7.0$ kOe, as an example. 
         \vspace{-3ex}}
\label{fig:fig1}
\end{figure}



\subsection{System description}
\label{sec:System descrpition}

The system we adopted in this study is schematically shown in Fig. \ref{fig:fig1}(a). 
The unit vectors pointing in the directions of the magnetization in the free and pinned layers are denoted as $\mathbf{m}$ and $\mathbf{p}$, respectively. 
We use a macrospin model throughout this paper.
The electric current flows in the $z$ direction, where a positive current corresponds to the electrons flowing from the free to pinned layer. 
The $x$ axis is parallel to the in-plane anisotropy axis in the free layer. 
The magnetization dynamics in the free layer is described by the LLG equation 
\begin{equation}
  \frac{d \mathbf{m}}{dt}
  =
  -\gamma
  \mathbf{m}
  \times
  \mathbf{H}
  -
  \gamma
  H_{\rm s}
  \mathbf{m}
  \times
  \left(
    \mathbf{p}
    \times
    \mathbf{m}
  \right)
  +
  \alpha
  \mathbf{m}
  \times
  \frac{d \mathbf{m}}{dt},
  \label{eq:LLG}
\end{equation}
where $\gamma$ and $\alpha$ are the gyromagnetic ratio and the Gilbert damping constant, respectively. 
The magnetic field consists of an in-plane anisotropy field $H_{\rm K}$, perpendicularly applied field $H_{\rm appl}$, and demagnetization field $-4\pi M$ as 
\begin{equation}
  \mathbf{H}
  =
  H_{\rm K}
  m_{x}
  \mathbf{e}_{x}
  +
  \left(
    H_{\rm appl}
    -
    4 \pi M 
    m_{z}
  \right)
  \mathbf{e}_{z},
  \label{eq:field}
\end{equation}
where $\mathbf{e}_{x}$ and $\mathbf{e}_{z}$ are the unit vectors pointing in the $x$ and $z$ directions.
For convention, we assume that the applied field points to the positive $z$ direction ($H_{\rm appl}>0$). 
Note that $H_{\rm appl}$ may include a stray field from the pinned layer. 
We assume that the demagnetization coefficient assigned to the perpendicular direction is one, for simplicity. 
For example, the demagnetization coefficient assigned to the perpendicular direction of the sample, 
the thickness and the diameter of which are 3 nm and and 85 nm, respectively, in Ref. \cite{hiramatsu16}, 
is calculated to be larger than 0.90 using the formula derived in Ref. \cite{tandon04}. 
The strength of the spin torque is given by \cite{slonczewski96,slonczewski05}
\begin{equation}
  H_{\rm s}
  =
  \frac{\hbar \eta j}{2e (1+\lambda \mathbf{m}\cdot\mathbf{p}) Md},
  \label{eq:H_s}
\end{equation}
where $j$, $d$, $\eta$, and $\lambda$ are the current density, thickness of the free layer, spin polarization of the current, and spin torque asymmetry, respectively. 
The spin torque has been calculated by using several theoretical models such as 
the interface scattering \cite{slonczewski96,brataas01}, 
the first-principles calculations \cite{stiles02,zwierzycki05}, 
the Boltzmann approach \cite{stiles02JAP,xiao04}, and the diffusive spin transport theory in bulk \cite{zhang02}. 
Although the characteristic parameters of the spin torque depend on the models, 
these theories basically deduce the same angular dependence of the spin torque. 
For example, $\eta$ in Ref. \cite{slonczewski05} is the spin polarization of the tunnel probability, 
and $\lambda=\eta^{2}$. 
The values of the parameters used in the following calculations are derived from a recent experiment \cite{hiramatsu16}, that is, 
$M=1300$ emu/cm${}^{3}$, $\gamma=1.764 \times 10^{7}$ rad/(Oe s), $d=3$ nm, $\eta=0.6$, $\lambda=\eta^{2}$, and $\alpha=0.01$. 
The in-plane anisotropy field is estimated to be 200 Oe \cite{comment1}. 


The magnetization dynamics described by the LLG equation can be regarded as a motion of a point particle on a sphere. 
It is useful to imagine an energy map of the free layer on this sphere because the self-oscillation occurs on a constant energy curve \cite{bertotti09text}. 
The energy density of the free layer is 
\begin{equation}
\begin{split}
  E
  &=
  -M 
  \int d 
  \mathbf{m}
  \cdot
  \mathbf{H}
\\
  &=
  -M 
  H_{\rm appl} 
  m_{z} 
  -
  \frac{MH_{\rm K}}{2}
  m_{x}^{2}
  +
  2\pi 
  M^{2}
  m_{z}^{2}. 
\end{split}
\end{equation}
Figure \ref{fig:fig1}(b) is an schematic view of the constant energy curves. 
The constant energy curves are classified into two regions. 
The curves rotating around an axis parallel to the $x$ axis represent energetically stable regions along the in-plane direction. 
On the other hand, the curves around the $z$ axis correspond to the orbit of an out-of-plane self-oscillation mentioned below. 
These two regions are separated by the constant energy curve of a saddle point energy. 




\begin{figure*}
\centerline{\includegraphics[width=2.0\columnwidth]{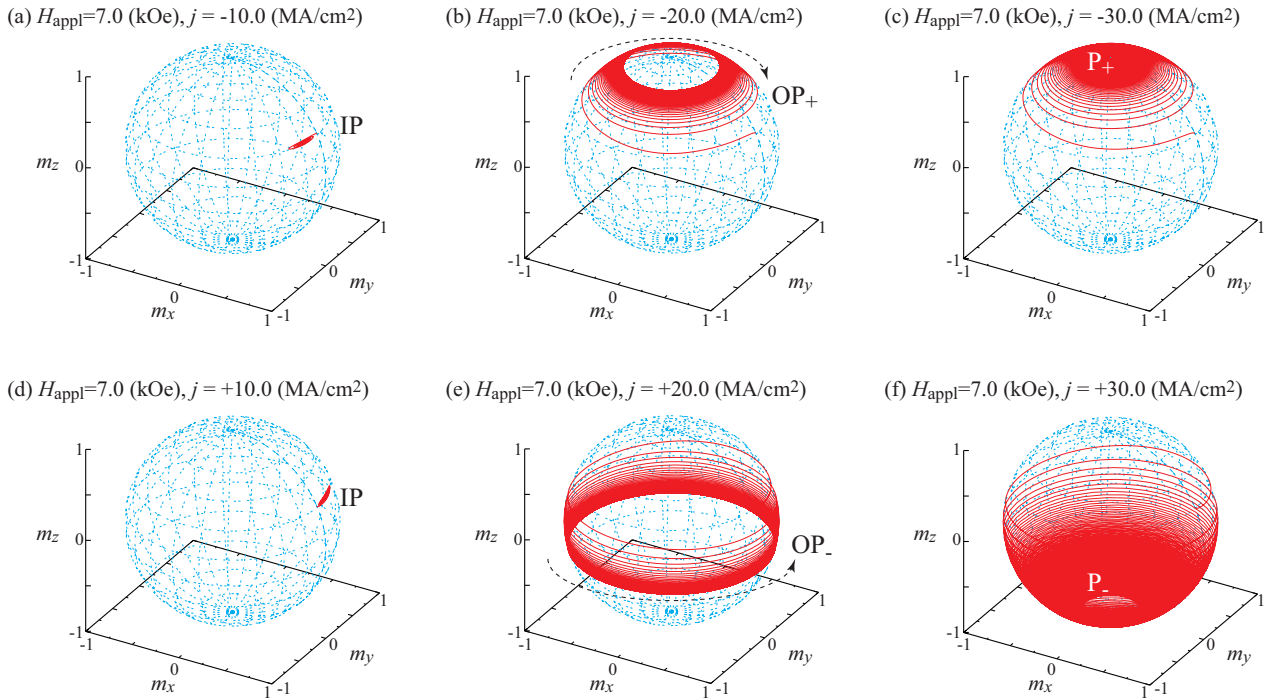}}
\caption{(Color online)
        Typical magnetization dynamics excited by negative currents obtained from numerical simulation of Eq. (\ref{eq:LLG}). 
        The magnitude of the applied field is set to be 7.0 kOe, whereas the currents are (a)-(c) negative and (d)-(f) positive. 
        The dotted lines in (b) and (e) indicate the oscillation directions, which are clockwise and counterclockwise, respectively, when viewing from the positive $z$ direction. 
        These dynamics are distinguished as (a), (d) IP, (b) OP${}_{+}$, (c) P${}_{+}$, (e) OP${}_{-}$, and (f) P${}_{-}$ states. 
         \vspace{-3ex}}
\label{fig:fig2}
\end{figure*}




\subsection{Dynamical phases}
\label{sec:Dynamical phases}

Figure \ref{fig:fig2} shows typical magnetization dynamics excited by the spin torque with negative current. 
The initial state is set to be in the energetically stable state. 
The currents are negative in Fig. \ref{fig:fig2}(a)-\ref{fig:fig2}(c). 
When the current magnitude is smaller than a threshold value, 
the damping torque overcomes the spin torque, and the magnetization stays in the in-plane region, 
as shown in Fig. \ref{fig:fig2}(a). 
When the current magnitude becomes larger than the threshold value, two possible dynamics appear. 
The first one is a self-oscillation around the $z$ axis shown in Fig. \ref{fig:fig2}(b), 
where the self-oscillation occurs on a constant energy curve shown in Fig. \ref{fig:fig1}(b). 
The other is moving to the perpendicular direction and eventually becomes parallel to the $z$ axis, as shown in Fig. \ref{fig:fig2}(c). 
Similar dynamics appears also by a positive current, 
as shown in Fig. \ref{fig:fig2}(d)-\ref{fig:fig2}(f), 
where the magnetization moves to a negative $z$ direction. 


For the sake of convention, we label these in-plane magnetized state, out-of-plane self-oscillation state, and perpendicularly magnetized state 
as IP, OP, and P, respectively in the following discussion; see Fig. \ref{fig:fig2}. 
Moreover, we add the subscript $\pm$ to the labels OP and P, such as OP${}_{\pm}$ and P${}_{\pm}$. 
For example, OP${}_{+(-)}$ means that the magnetization moves to the positive (negative) $z$ direction by a negative (positive) current and shows a self-oscillation. 
Notice here that the chirality (rotation direction) of the oscillation, viewing from the positive $z$ direction,
in the OP${}_{+}$ state is clockwise whereas that in the OP${}_{-}$ state is counterclockwise, 
as shown in Fig. \ref{fig:fig2}(b) and \ref{fig:fig2}(e), respectively. 



\subsection{Relation to MAMR}
\label{sec:Relation to MAMR}

In this section, let us briefly discuss the relation between the oscillation properties of the STO found in Sect. \ref{sec:Dynamical phases} and MAMR. 



\begin{figure}
\centerline{\includegraphics[width=1.0\columnwidth]{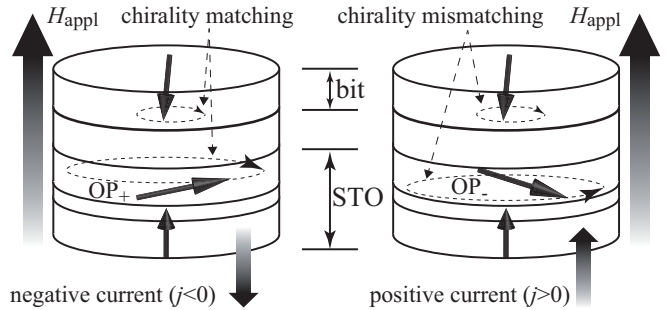}}
\caption{(Color online)
         Schematic illustration of possible situations occurred in a nanopillar structure used in experiments [\onlinecite{suto12,hiramatsu16}] 
         for a demonstration of MAMR by microwaves emitted from an STO. 
         A negative current excites a self-oscillation in the STO having the chirality of clockwise, 
         whereas a positive current excites a self-oscillation having the chirality of counterclockwise, 
         which correspond to the OP${}_{+}$ and OP${}_{-}$ states, respectively. 
         The chirality of a bit is clockwise. 
         The chirality matching between the STO and bit is achieved for a negative current case. 
         \vspace{-3ex}}
\label{fig:fig3}
\end{figure}



As written in the Introduction, 
the recording bit will be aligned parallel to the current in the commercial design for high density media. 
According to the experiments related to MAMR \cite{suto12,hiramatsu16}, however, 
we assume that a perpendicularly magnetized ferromagnet corresponding to a recording bit is placed onto the STO, 
as schematically shown in Fig. \ref{fig:fig3}.
In these experiments, the applied field $H_{\rm appl}$ plays a role of the reversal field of the recording bit, which also acts on the STO. 
Figure \ref{fig:fig3} schematically shows possible situations occurring in this nanopillar structure. 
When a negative current is injected, the magnetization in the STO shows a self-oscillation with the chirality of clockwise (OP${}_{+}$ state), as shown in Fig. \ref{fig:fig2}(b). 
On the other hand, when a positive current is injected, a self-oscillation with the chirality of counterclockwise (OP${}_{-}$ state) is excited in the STO. 


The principle of MAMR is that the microwave assists the magnetization reversal 
when the chirality of the microwave is identical to that of the magnetization in the bit \cite{kudo15}. 
When the magnetic field is applied in the positive $z$ direction, 
we will assume that the magnetization in the bit is initially pointing in the negative $z$ direction, 
as shown in Fig. \ref{fig:fig3}. 
In this case, the chirality of the magnetization in the bit is clockwise. 
Therefore, the OP${}_{+}$ state excited by a negative current should be used as a microwave source for MAMR due to chirality matching; see Fig. \ref{fig:fig3}. 
Accordingly, in the following, we principally focus on the magnetization dynamics excited by the negative current. 


\subsection{Theoretical formulas of characteristic currents}
\label{sec:Theoretical formulas of characteristic currents}

Let us call the characteristic current determining the boundary between the IP and the other states as threshold current. 
We extend a theory developed in our previous work \cite{taniguchi16} to be applicable to the present system. 
In Ref. \cite{taniguchi16}, an instability condition of an STO was investigated, 
where the STO does not have an in-plane anisotropy and an applied field points to an arbitrary direction. 
The present system, on the other hand, has an in-plane anisotropy, and an applied field points to the perpendicular direction. 
Two mechanisms destabilizing the IP state exist in this type of STO. 
In the first mechanism, the magnetization absorbs energy from the work done by spin torque 
during the time shorter than a precession period around the in-plane axis, 
and overcomes the potential barrier \cite{taniguchi16}. 
The threshold current in this mechanism is given by 
(the derivation is summarized in Appendix \ref{sec:AppendixA})
\begin{equation}
\begin{split}
  j_{{\rm th}\pm}
  =&
  \frac{2 e Md}{\hbar \eta}
  4\pi M 
  \frac{\Delta \varepsilon+\alpha(1-h^{2}+\kappa)I_{\alpha \pm}}{I_{{\rm s}\pm}},
  \label{eq:jth}
\end{split}
\end{equation}
where $h=H_{\rm appl}/(4\pi M)$ and $\kappa=H_{\rm K}/(4\pi M)$. 
The subscript $\pm$ indicates that $j_{{\rm th}+(-)}$ is the current necessary to move the magnetization to the positive (negative) $z$ direction. 
In our definition of the current direction, $j_{{\rm th}+(-)}$ becomes negative (positive). 
The dimensionless potential barrier, 
\begin{equation}
  \Delta 
  \varepsilon
  =
  \frac{\kappa}{2}
  \left(
    1
    -
    \frac{h^{2}}{1+\kappa}
  \right), 
\end{equation}
is the energy difference between the stable state and the saddle point of $E$. 
We also introduce $I_{\alpha \pm}$ and $I_{{\rm s}\pm}$, which are proportional to the dissipation due to damping torque and work done by spin torque. 
They are given by 
\begin{equation}
  I_{\alpha \pm}
  =
  \frac{\sqrt{\kappa(1-h^{2})}}{1+\kappa}
  \mp
  \frac{\kappa h   \cos^{-1} \left[ \frac{\pm \kappa h}{\sqrt{\kappa (1-h^{2}+\kappa)}} \right]}
    {(1+\kappa)^{3/2}},
  \label{eq:I_alpha}
\end{equation}
\begin{equation}
\begin{split}
  &
  I_{{\rm s}\pm}
  =
  \frac{\pm h   \cos^{-1} \left[ \frac{\pm\kappa h}{\sqrt{\kappa(1-h^{2}+\kappa)}} \right]}
    {\lambda \sqrt{1+\kappa}}
\\
  &\ \ \ \ \mp
  \frac{(h+\lambda)  \cos^{-1}\left[ \frac{\pm \kappa(h+\lambda)}{(1+\lambda h) \sqrt{\kappa(1-h^{2}+\kappa)}} \right]}
    {\lambda \sqrt{(1+\lambda h)^{2}+\kappa (1-\lambda^{2})}}.
  \label{eq:I_s}
\end{split}
\end{equation}
We should note here that, in this mechanism, either a positive or negative current can destabilize the magnetization. 
This is because, if we focus on the magnetization dynamics shorter than a precession period, 
spin torque excited by either a positive or negative current can inject finite energy to the STO to overcome the dissipation due to the damping torque; see Appendix \ref{sec:AppendixA}. 


Another mechanism is slow excitation of the magnetization dynamics, 
where the magnetization precesses around the IP state and gradually increases its amplitude. 
The threshold current density due to this mechanism is obtained from the linearized LLG equation \cite{sun00,grollier03,morise05}, 
and is given by (see also Appendix \ref{sec:AppendixA}) 
\begin{equation}
  j_{\rm c}
  =
  \frac{2 \alpha eMd}{\hbar \eta \tilde{\lambda}}
  4\pi M 
  \left[
    \frac{1}{2}
    +
    \kappa
    -
    \frac{h^{2}}{2(1+\kappa)}
  \right],
  \label{eq:jc}
\end{equation}
where we define 
\begin{equation}
  \tilde{\lambda}
  =
  \frac{p_{Z}+[\Lambda(1-p_{z}^{2})/2]}{1+\lambda p_{Z}}, 
\end{equation}
\begin{equation}
  \Lambda
  =
  \frac{\lambda}{1+\lambda p_{Z}}, 
\end{equation}
\begin{equation}
  p_{Z}
  =
  \frac{h}{1+\kappa}. 
\end{equation}
Note here that $j_{\rm c}$ is a positive quantity. 
This is because the spin torque averaged over a precession period points to the antiparallel direction to the averaged damping torque 
only when the positive current is injected; 
when a negative current is injected, on the other hand, the averaged torque points to the parallel direction to the damping torque. 


Practically, the instability of the magnetization occurs when the current becomes larger than one of the two characteristic currents \cite{taniguchi16}. 
This means that the threshold current is given by ${\rm min}[j_{{\rm th}\pm},j_{\rm c}]$. 
Therefore, the threshold current for a positive current is given by ${\rm min}[j_{{\rm th}-},j_{\rm c}]$, 
whereas that for a negative current is $j_{{\rm th}+}$. 
A similar result is obtained for a different system \cite{taniguchi16APEX}. 
We also introduce another characteristic current necessary to move the magnetization parallel to the $z$ direction given by 
\begin{equation}
  j_{{\rm u}\pm}
  =
  \mp
  \frac{2 \alpha e (1 \pm \lambda)Md}{\hbar \eta}
  4\pi M 
  \left(
    1
    \mp
    h
    +
    \frac{\kappa}{2}
  \right).
  \label{eq:ju}
\end{equation}


Let us explain the physical meanings of these characteristic currents 
by focusing on the case of the negative current. 
We should remind the readers that $j_{{\rm th}+}$ is a current necessary to move the magnetization from the IP state, 
whereas $j_{{\rm u}+}$ is a current necessary to move the magnetization to the P${}_{+}$ state. 
When the current magnitude is smaller than $|j_{{\rm th}+}|$, 
the spin torque cannot overcome the damping torque, and thus, the IP state is stable. 
When the current magnitude is larger than the threshold value $|j_{{\rm th}+}|$, on the other hand, the IP state is destabilized by the spin torque, 
and the magnetization moves to a positive $z$ direction. 
The magnetization dynamics after destabilizing the IP state depends on the magnitudes of $j_{{\rm th}+}$ and $j_{{\rm u}+}$. 
When $j_{{\rm th}+}/j_{{\rm u}+}<1$, the OP${}_{+}$ (an out-of-plane self-oscillation) state appears 
on the condition that the current density is in the range of $|j_{{\rm th}+}|<|j|<|j_{{\rm u}+}|$. 
In this case, the spin torque balances the damping torque at a certain out-of-plane self-oscillation state. 
If the current magnitude $|j|$ becomes larger than $|j_{{\rm u}+}|$, the magnetization moves to the P${}_{+}$ state. 
When $j_{{\rm th}+}/j_{{\rm u}+}>1$, however, 
the magnetization directly moves to the P${}_{+}$ without showing self-oscillation. 
In this case, the OP${}_{+}$ state does not appear for any value of the current. 


An important point is that the magnitudes of $j_{{\rm th}+}$ and $j_{{\rm u}+}$ depend on the value of the applied field. 
Notice that the applied field $H_{\rm appl}(>0)$ points to the opposite direction to the demagnetization field $-4\pi Mm_{z}$, 
where $m_{z}>0$ since we are interested in the magnetization dynamics excited by a negative current. 
When the magnitude of the applied field is relatively small, $j_{{\rm th}+}/j_{{\rm u}+}<1$. 
This is because $j_{{\rm u}+}$ is principally determined by the effective field in the perpendicular direction 
whereas $j_{{\rm th}+}$ by the in-plane anisotropy field, 
and the demagnetization field is usually much larger than the in-plane anisotropy field. 
However, the condition $j_{{\rm th}+}/j_{{\rm u}+}<1$ is no longer satisfied when the applied field becomes larger than a critical value $H_{\rm c}$. 
This is because a large field significantly suppresses the demagnetization field, and therefore, results in small $j_{{\rm u}+}$. 



In summary, the magnitude relation between $j_{{\rm th}+}$ and $j_{{\rm u}+}$ 
changes from $j_{{\rm th}+}/j_{{\rm u}+}<1$ to $j_{{\rm th}+}/j_{{\rm u}+}>1$ 
when $H_{\rm appl}$ becomes larger than a critical value $H_{\rm c}$. 
Consequently, according to the above discussion, the OP${}_{+}$ state disappears. 
These arguments will be confirmed in the following sections. 
It should be noted that we assumed small value for the parameter $\lambda$ in the above discussion. 
The role of the parameter $\lambda$ is briefly summarized in Appendix \ref{sec:AppendixA}.





\begin{figure*}
\centerline{\includegraphics[width=2.0\columnwidth]{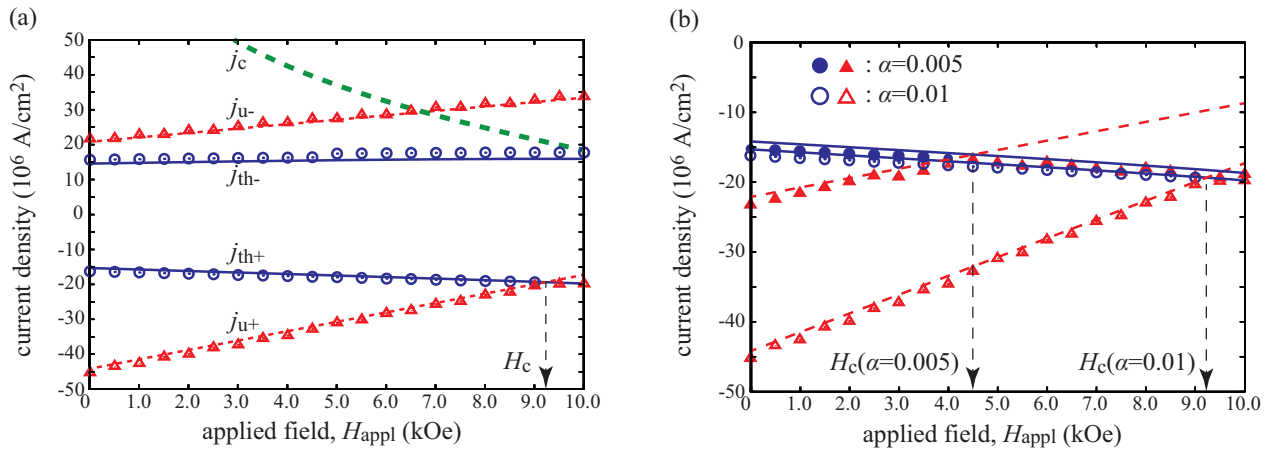}}
\caption{(Color online)
        (a) Dependences of $j_{{\rm th}\pm}$ (solid line), $j_{{\rm u}\pm}$ (dotted line), and $j_{\rm c}$ (dashed line) on the applied field $H_{\rm appl}$. 
            The critical field corresponding to a crossing point of $j_{{\rm th}+}$ and $j_{{\rm u}+}$ is denoted as $H_{\rm c}$. 
            The circles are the numerically evaluated currents above which an out-of-plane self-oscillation is excited. 
            On the other hand, the triangles are the numerically evaluated currents above which the magnetization becomes parallel to the $z$ axis. 
            The self-oscillation appears when the former current is smaller than the latter. 
        (b) The solid circles and triangles respectively indicate the numerically evaluated current to excite an out-of-plane self-oscillation and to move the magnetization parallel to the $z$ axis 
            for $\alpha=0.005$ in the case of negative currents. 
            For comparison, these currents for $\alpha=0.01$, which are identical to those in (a), are also shown by the open circles and triangles. 
            The theoretical formulas of $j_{{\rm th}+}$ and $j_{{\rm u}+}$ are also shown by the solid and dotted lines, respectively. 
         }
\label{fig:fig4}
\end{figure*}




\subsection{Critical field}
\label{sec:Critical field}

In this section, we confirm the above argument that the OP${}_{+}$ state disappears by increasing the magnitude of the applied magnetic field. 


Figure \ref{fig:fig4}(a) shows the dependences of the theoretical values of the characteristic currents, $j_{{\rm th}\pm}$, $j_{{\rm u}\pm}$, and $j_{\rm c}$, 
on the applied field $H_{\rm appl}$ by the solid, dotted, and dashed lines, respectively. 
As mentioned above, the magnitude relation between $j_{{\rm th}+}$ and $j_{{\rm u}+}$ 
changes from $j_{{\rm th}+}/j_{{\rm u}+}<1$ to $j_{{\rm th}+}/j_{{\rm u}+}>1$ 
when $H_{\rm appl}$ becomes larger than a certain value $H_{\rm c}$. 
For the present parameters, $H_{\rm c}$ is 9.23 kOe, 
which is much smaller than the demagnetization field $4\pi M \simeq 16.3$ kOe. 


We study the validity of the theoretical formula of the threshold current 
by comparing with the numerical simulation of the LLG equation at zero temperature. 
In the numerical simulation, we calculate the magnetization dynamics in the time range of $0 \le t \le 1.0$ $\mu$s 
with the time step of $\Delta t=0.2$ ps. 
The initial state is the energetically stable state, i.e., $\mathbf{m}(t=0)=\mathbf{m}_{\rm stable}$. 
We define the numerically evaluated threshold current as the minimum current satisfying a condition 
$|m_{x}(t=t_{\rm max})-m_{x}(t=t_{\rm max}-\Delta t)|>10^{-10}$, where $t_{\rm max}=1.0$ $\mu$s. 
This condition indicates that the magnetization is in an oscillation state. 
The circles in Fig. \ref{fig:fig4}(a) show the numerically evaluated value of the threshold current determined using this condition. 
As shown, a good agreement with the theoretical formula is obtained. 
The small difference between the theoretical formula and the numerical simulation is caused by the approximation from Eq. (\ref{eq:energy_equation_orig}) to (\ref{eq:energy_equation}), which is used in the derivation of Eq. (\ref{eq:jth})
We also study the minimum current necessary to move the magnetization parallel to the $z$ axis. 
This current is numerically evaluated from conditions that $|m_{z}(t=t_{\rm max})|>0.99$ and $|m_{x}(t=t_{\rm max})-m_{x}(t=t_{\rm max}-\Delta t)|<10^{-10}$. 
These conditions corresponds to the case where the magnetization becomes nearly parallel to the $z$ axis and stops its dynamics. 
The triangles in Fig. \ref{fig:fig4}(a) are the numerically evaluated value of the current necessary to move the magnetization to the $z$ direction, 
which also show good agreement with the theoretical formula of $j_{{\rm u}\pm}$. 
Accordingly, the numerically evaluated critical field, indicated as the crossing point of the circles and triangles, also shows good agreement with 
that estimated from the theoretical formulas, corresponding to the crossing point of the solid and dotted line in Fig. \ref{fig:fig4}(a). 
Note here that the numerically evaluated threshold current for the negative current case does not exist when $H_{\rm appl}>H_{\rm c}$. 
This is because the magnetization directly moves to the $z$ direction without showing self-oscillation in the OP${}_{+}$ state. 


We should remind the readers that the results in Fig. \ref{fig:fig4}(a) are obtained for $\alpha=0.01$. 
In Fig. \ref{fig:fig4}(b), the solid circles and triangles respectively indicate 
the numerically evaluated currents to excite the self-oscillation and to move the magnetization parallel to the $z$ axis 
for $\alpha=0.005$ in the case of negative current. 
For comparison, the currents obtained in Fig. \ref{fig:fig4}(a) shown by the open circles and triangles are also shown in Fig. \ref{fig:fig4}(b). 
As shown in the figure, the critical field for $\alpha=0.005$, which corresponds to the crossing point of the solid circles and triangles, is about 4.51 kOe. 
This value is much smaller than that for $\alpha=0.01$, which is about 9.23 kOe, as mentioned above. 
The reason why the critical field decreases with decreasing $\alpha$ is that 
the value of $j_{{\rm th}\pm}$ is dominantly determined by a term independent of the damping in Eq. (\ref{eq:jth}), 
whereas $j_{{\rm u}\pm}$ is proportional to $\alpha$. 
Therefore, by reducing the value of $\alpha$, $j_{{\rm u}\pm}$ crosses $j_{{\rm th}\pm}$ at a small field, 
resulting in the decrease of the critical field $H_{\rm c}$. 


In summary, the numerical simulation at zero temperature confirms the validity of the theoretical formulas characterizing the magnetization dynamics. 
An out-of-plane self-oscillation state disappears when the magnitude of the applied field becomes larger than a critical value. 
The value of the critical field decreases with the decrease in the damping constant. 




\begin{figure}
\centerline{\includegraphics[width=1.0\columnwidth]{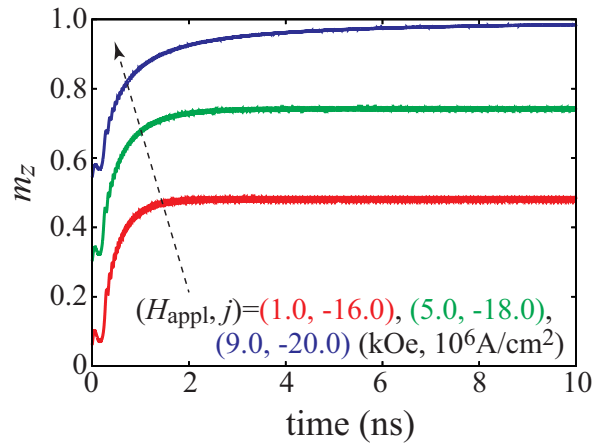}}
\caption{(Color online)
         Time evolutions of $m_{z}$ for $H_{\rm appl}=1.0$ (red), $5.0$ (green), and $9.0$ (blue) kOe. 
         The current densities, $j=-16.0$, $-18.0$, and $-20.0$ MA/cm${}^{2}$ are close to $j_{\rm th+}$ for each field value.
         \vspace{-3ex}}
\label{fig:fig5}
\end{figure}




\subsection{Applicability of present formula}
\label{sec:Applicability of present formula}

In the derivation of Eq. (\ref{eq:jth}), we assume that the electric current is injected from $t=0$ in a step-function manner, 
in accordance with the experimental conditions mentioned in Ref. \cite{hiramatsu16}. 
Before the initial time $t=0$, the current is absent, and therefore, 
the magnetization initially stays at the energetically stable point located in the $xz$ plane; see Appendix \ref{sec:AppendixA}. 

There is another possible situation. 
Note that the spin torque in the present system is finite at the energetically stable state. 
Therefore, when the current magnitude is small, the magnetization will move to a steady point where the sum of all torques become zero. 
For example, Ref. \cite{ebels08} investigates the stability condition of the spin torque oscillator around the steady point. 
This formulation is applicable when the current is injected slowly to the STO, 
and therefore, the magnetization can move from the energetically stable state to the steady point before showing a self-oscillation. 
In fact, the numerical simulation of Ref. \cite{ebels08} allows the magnetization to move from the initial energetically stable state to the steady point 
by using a current with a certain ramp of a few nanoseconds 
\cite{comment_ebels}. 

We should note that the ramp time in experiments is finite. 
The assumption that the current is injected in a step-function manner works well 
when the ramp time of the current is shorter than the time necessary to move from the IP to OP or P state. 
Figure \ref{fig:fig5} shows the time evolutions of $m_{z}$ for several values of the applied field, 
$H_{\rm appl}=1.0$, $5.0$, and $9.0$ kOe. 
The current densities, $j=-16.0$, $-18.0$, and $-20.0$ MA/cm${}^{2}$, for each field are chosen to be close to the threshold value $j_{{\rm th}+}$. 
Note that $m_{z}$ becomes almost constant when the magnetization moves from the IP to OP or P state. 
As can be seen from Fig. \ref{fig:fig5}, the time necessary to move from the IP to OP or P state is typically on the order of nanoseconds. 
Therefore, the present formula works well when the ramp time is less than a few nanoseconds. 

We would like to emphasize here that both the previous work, such as Ref. \cite{ebels08}, 
and our present work are essential for investigating the instability conditions of the spin torque oscillators. 
The analysis in Ref. \cite{ebels08} can be a powerful tool to analyze the magnetization dynamics in a situation where an initial condition is close to the steady point. 
On the other hand, the present work provides the theoretical formula of the instability condition 
when the magnetization initially locates near the energetically stable state, and where the injected current abruptly jumps from zero to a finite value. 
This approach will be useful in MAMR using microwaves generated from an STO, 
where a fast magnetization reversal process shorter than a few nanoseconds in a recording bit caused by microwaves generated from an STO is required. 
In this case, a current which is almost in a step-function manner is necessary \cite{kudo15}. 
We should also note here that a similar situation, i.e., the instability threshold depends on the initial state, was found in a different problem; see Appendix \ref{sec:AppendixB}.



\section{Numerical simulation at finite temperature}
\label{sec:Numerical simulation at finite temperature}

The purposes of this section are as follows. 
First, we confirm that the results shown in Sect. \ref{sec:System description and LLG equation}, 
which are obtained at zero temperature, are also valid at finite temperature. 
Second, we study the relation between the current and oscillation frequency of the STO. 
This is because the current-frequency relation has been often measured in experiments \cite{hiramatsu16}.



\begin{figure*}
\centerline{\includegraphics[width=2.0\columnwidth]{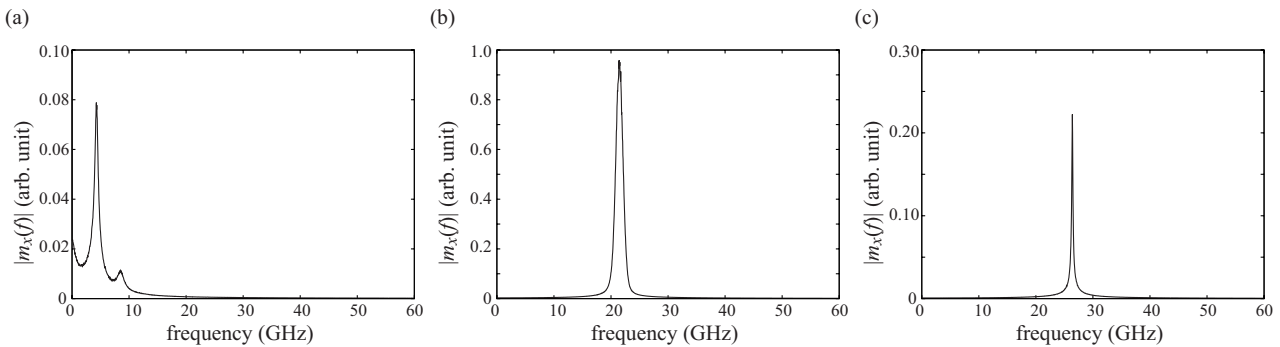}}
\caption{(Color online)
        The Fourier transformation of $m_{x}(t)$ at finite temperature, $T=300$ K. 
        The magnitude of the applied field $H_{\rm appl}$ is $7.0$ kOe, whereas the current densities are 
        (a) $-10$, (b) $-20$, and (c) $-30$ MA/cm${}^{2}$. 
         \vspace{-3ex}}
\label{fig:fig6}
\end{figure*}



\subsection{Temperature effect}
\label{sec:Temperature effect}

The temperature effect is taken into account by adding random torque, 
$-\gamma \mathbf{m}\times\mathbf{h}$, originated from thermal fluctuation to the right hand side of Eq. (\ref{eq:LLG}). 
The components of the random field $h_{k}(t)$ ($k=x,y,z$) satisfy the fluctuation-dissipation theorem \cite{brown63} 
\begin{equation}
  \langle 
    h_{k}(t)
    h_{\ell}(t^{\prime})
  \rangle 
  =
  \frac{2 \alpha k_{\rm B}T}{\gamma MV}
  \delta_{k \ell}
  \delta(t-t^{\prime}).
  \label{eq:FDT}
\end{equation}
In the numerical simulation, the random field is given by \cite{russek05}
\begin{equation}
  h_{a}(t)
  =
  \sqrt{
    \frac{2\alpha k_{\rm B}T}{\gamma MV \Delta t}
  }
  \xi_{a}(t).
\end{equation}
The value of the volume $V$ of the free layer is determined from the experiment \cite{hiramatsu16} as $V=\pi r^{2}d$ with the radius $r=40$ nm, 
whereas the temperature is $T=300$ K. 
White noise $\xi$ is defined from two random numbers, $\zeta_{a}$ and $\zeta_{b}$, in the range of $0 < \zeta_{a},\zeta_{b} \le 1$ by the Box-Muller transformation as 
$\xi_{a}=\sqrt{-2 \log \zeta_{a}} \sin(2 \pi \zeta_{b})$ and $\xi_{b}=\sqrt{-2 \log \zeta_{a}} \cos(2 \pi \zeta_{b})$. 
We calculate the magnetization dynamics from the LLG equation with the white noise 1000 times for a given set of the applied field and the current density. 
We pick up $N_{\rm FFT}=2^{17}=131072$ data of $\mathbf{m}(t)$ for the time $t$ in the range of $t_{\rm max}-N_{\rm FFT}\Delta t<t \le t_{\rm max}$. 
The Fourier transformation is applied to the sampled $m_{x}(t)$. 


Figures \ref{fig:fig6}(a)-(c) show examples of the spectra $|m_{x}(f)|$ at $H_{\rm appl}=7.0$ kOe, 
where the values of the current density are (a) $-10$, (b) $-20$, and (c) $-30$ MA/cm${}^{2}$. 
We note that the magnetization dynamics at these currents correspond to the IP, OP${}_{+}$, and P${}_{+}$ states 
shown in Figs. \ref{fig:fig2}(a)-\ref{fig:fig2}(c), respectively. 
At the IP${}_{+}$ state, the magnetization cannot move from an energetically stable region around an in-plane axis. 
Thus, the magnetization immediately moves to the final state, and stop its dynamics. 
As a result, the dominant component of $|m_{x}(f)|$ appears at $f=0$. 
Other than this offset, a peak appears at $4.3$ GHz, corresponding to the magnoise frequency \cite{smith09,tamaru17}, as shown in Fig. \ref{fig:fig6}(a). 
Since the magnetization oscillates around an axis closely parallel to the $x$ axis, as shown in Fig. \ref{fig:fig2}(a), 
another peak appears at the second harmonic. 
At the OP${}_{+}$ state, an out-of-plane self-oscillation with a large amplitude of $m_{x}$ is excited. 
Thus, a large peak appears at the self-oscillation frequency, as shown in Fig. \ref{fig:fig6}(b). 
The magnoise also appears at the P${}_{+}$ state, as shown in Fig. \ref{fig:fig6}(c). 
In this case, the power of $|m_{x}(f)|$ is relatively small compared with that shown in Fig. \ref{fig:fig6}(b) 
because the magnetization points to the $z$ direction, and thus, the amplitude of $m_{x}(t)$ is small. 



\begin{figure*}
\centerline{\includegraphics[width=2.0\columnwidth]{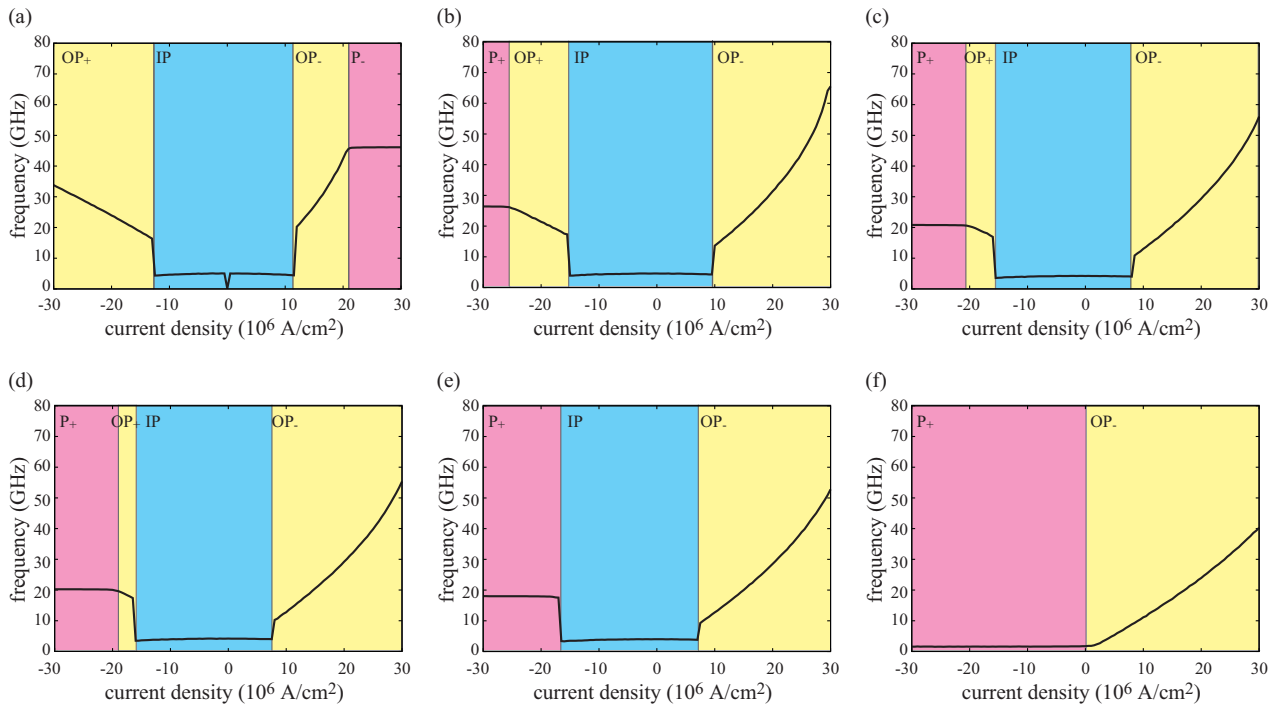}}
\caption{(Color online)
         Relations between the current density and oscillation frequency for the perpendicular field of (a) 0, (b) 7.0, (c) 9.0, (d) 9.2, (e) 10.0, and (f) 17.0 kOe. 
         Random torque due to thermal fluctuation at $T=300$ K is added. 
         The labels, IP, OP${}_{\pm}$, and P${}_{\pm}$, distinguish the dynamical phase of the magnetization. 
         Remember that the OP${}_{+}$ state is necessary for MAMR. 
         \vspace{-3ex}}
\label{fig:fig7}
\end{figure*}




\subsection{Current-frequency relation}
\label{sec:Current-frequency relation}

Here, we confirm that the results obtained in Sect. \ref{sec:System description and LLG equation} is still valid at finite temperature from the current-frequency relation. 
We note here that the thermal effect is practically necessary to distinguish the IP and P${}_{\pm}$ states from the current-frequency relation. 
This is because the magnetization dynamics at zero temperature in both cases eventually terminated, as shown in Figs. \ref{fig:fig2}(a) and \ref{fig:fig2}(b), 
and thus, the frequencies become zero. 
At finite temperature, however, the IP and P${}_{\pm}$ can be distinguished from the magnoise frequency. 
The oscillation frequency evaluated from the numerical simulation corresponds to the lowest frequency, except offset, 
giving a local maximum of the Fourier spectrum $|m_{x}(f)|$. 
We also note that the thermal fluctuation makes the definitions of characteristic currents ambiguous and different from those defined at zero temperature. 


As mentioned in Sect. \ref{sec:Relation to MAMR}, 
the OP${}_{+}$ state is necessary for MAMR because of the chirality matching between the STO and a recording bit. 
Therefore, let us study how the OP${}_{+}$ at finite temperature changes by changing the magnitude of the applied field. 


Figures \ref{fig:fig7}(a)-\ref{fig:fig7}(f) summarize the current-frequency relations for several values of the applied magnetic field. 
When the applied field is sufficiently small, 
we observe all of the IP, OP${}_{\pm}$, and P${}_{\pm}$ states, as shown in Figs. \ref{fig:fig7}(a) and \ref{fig:fig7}(b). 
However, the range of the OP${}_{+}$ state becomes narrow with increasing $H_{\rm appl}$. 
When the magnitude of the applied field becomes larger than a critical value, 
the OP${}_{+}$ state disappears, as can be seen from Figs. \ref{fig:fig7}(c), \ref{fig:fig7}(d), and \ref{fig:fig7}(e). 
This result is consistent with the result obtained at zero temperature, 
and indicates the existence of the critical field even at finite temperature. 
We note that it is difficult to distinguish the OP${}_{+}$ state close to $\mathbf{m}\simeq \mathbf{e}_{z}$ and P${}_{+}$ state at finite temperature 
because the stochastic oscillation around the $z$ axis due to the thermal fluctuation makes the boundary between these states ambiguous. 
As a result, an evaluation of the critical field becomes relatively unclear. 
Nevertheless, the critical field estimated from Fig. \ref{fig:fig7}, which is in the range between 9.0 and 10.0 kOe, 
is also consistent with that obtained at zero temperature, $H_{\rm c}=9.23$ kOe.  
Similarly, the instability thresholds from the IP state to OP${}_{+}$ or P${}_{+}$ state in Fig. \ref{fig:fig7}(a)-\ref{fig:fig7}(e) 
show good agreement with those obtained from the zero-temperature calculation in Fig. \ref{fig:fig4}(b). 
For example, the currents corresponding to the boundary between the IP and OP${}_{+}$ and P${}_{+}$ states in Fig. \ref{fig:fig7}(a)-\ref{fig:fig7}(e) 
are (-13.0,-15.5,-16.0,-16.5,-17.0) $\times 10^{6}$ A/cm${}^{2}$
for $H_{\rm appl}=(0,7.0,9.0,9.2,10.0)$ kOe, 
whereas those estimated from Fig. \ref{fig:fig4} are (-15.3,-18.3,-19.2,-19.3,-19.8) $\times 10^{6}$ A/cm${}^{2}$. 
We should note that the instability threshold at the finite temperature is lower than that at the zero temperature 
because the thermal fluctuation excites an oscillation of the magnetization around the energetically stable initial state, 
and therefore, transition from IP or P state becomes easy. 
When the applied field is in the range of $H_{\rm c}<H_{\rm appl}\lesssim 4\pi M$, 
the magnetization directly moves from the IP to the perpendicularly magnetized state (P${}_{+}$) without showing the out-of-plane self-oscillation. 
When the field magnitude becomes sufficiently larger than the demagnetization field, 
the magnetization initially points to the positive $z$ direction, and therefore, the IP state disappears, as shown in Fig. \ref{fig:fig7}(f). 
On the other hand, the range of the OP${}_{-}$ state becomes wide with increasing $H_{\rm appl}$. 
This state is, however, not applicable to MAMR because of the mismatching of the chirality; see Fig. \ref{fig:fig3}. 


In summary, the critical field $H_{\rm c}$ above which the out-of-plane self-oscillation state disappears exists even at finite temperature. 
Although the evaluation of $H_{\rm c}$ at finite temperature becomes relatively unclear due to the thermal fluctuation, 
its value is basically consistent with that obtained at zero temperature.

\section{Conditions to observe MAMR}
\label{sec:Conditions to observe MAMR}


\subsection{In nanopillar devices}

Let us first consider the case of nanopillar-structured STO. 
This is one of the suitable case- studies in which the theory developed in this paper can be applied.
As mentioned in Sect. \ref{sec:Relation to MAMR}, 
the OP${}_{+}$ state should be excited when the type of STO in this study is used as a microwave generator for MAMR. 
However, the OP${}_{+}$ state does not appear when $H_{\rm appl}$ becomes larger than the critical value $H_{\rm c}$, as shown in the above sections. 
This fact implies that the reversal field of the bit in MAMR should be smaller than $H_{\rm c}$ of the STO. 
This becomes a strong constraint to observe MAMR in the experimental geometry which is currently used \cite{suto12,hiramatsu16}. 



The value of $H_{\rm c}$ can be theoretically evaluated from the condition,
\begin{equation}
  \frac{j_{{\rm th}+}}{j_{{\rm u}+}}
  =
  1. 
  \label{eq:condition_Hc}
\end{equation}
The theoretical values of $j_{{\rm th}+}$ shown in Fig. \ref{fig:fig4}(b) indicate that 
$j_{{\rm th}\pm}$ is principally determined by a term independent of the damping constant. 
Then, we note that the field magnitude satisfying Eq. (\ref{eq:condition_Hc}) decreases with decreasing $\alpha$ 
because a damping-independent term is the dominant factor in determining $j_{{\rm th}\pm}$ whereas $j_{{\rm u}\pm}$ is proportional to the damping. 
Therefore, a small damping is not preferable to achieve a wide range of the OP${}_{+}$, 
although a material with a small damping is of interest from the view point of the power reduction of an STO. 
An analytical and approximated formula of $H_{\rm c}$ is studied in Appendix \ref{sec:AppendixC}. 
On the other hand, when the damping constant is relatively large, $\alpha \gtrsim \sqrt{\kappa}$, 
the threshold current $j_{{\rm th}\pm}$ is dominated by the damping-dependent term; see Eq. (\ref{eq:jth_limit}) in Appendix \ref{sec:AppendixA}. 
In this case, the critical field will be independent of the damping constant. 


We also note that $j_{{\rm th}+}=0$ in the absence of the in-plane anisotropy field. 
Thus, $H_{\rm c}$ does not appear when $H_{\rm K}=0$, 
and in this case, the OP${}_{+}$ state always appears for the field $H_{\rm appl}$ smaller than the demagnetization field $4\pi M$. 
Therefore, the reduction of the in-plane anisotropy, 
which originates from shape design \cite{houssameddine07} and/or fabrication process \cite{comment1}, 
will be a critical issue for the further development of MAMR. 


There are other conditions to observe MAMR. 
For example, the microwave frequency emitted from an STO should be less than the critical frequency \cite{taniguchi14,taniguchi16PRB} 
\begin{equation}
  f_{\rm MAMR}
  =
  \frac{\gamma H_{\rm K(b)}}{2\pi}
  \frac{(H_{\rm ac}/H_{\rm K(b)})^{2/3}}{\sqrt{1-(H_{\rm ac}/H_{\rm K(b)})^{2/3}}}
  \left[
    2
    -
    \frac{5}{3}
    \left(
      \frac{H_{\rm ac}}{H_{\rm K(b)}}
    \right)^{2/3}
  \right],
\end{equation}
where $H_{\rm K(b)}$ is the perpendicular anisotropy field of a recording bit, 
whereas $H_{\rm ac}$ is the microwave amplitude from the STO, which depends on the distance between the STO and the bit. 
A further discussion is, however, beyond the scope of this paper. 


\subsection{Design for commercial use}

In the previous section, we have considered a nanopillar structure used in the experiments \cite{suto12,hiramatsu16}. 
This assumption implies that the reversal field of the ferromagnet corresponding to a recording bit also acts on an STO. 
As emphasized above, the condition to excite a self-oscillation in the STO leads to a restriction on the reversal field of the recording bit in the nanopillar structure. 
In designing MAMR for commercial use, on the other hand, the STO is separated from the recording bit. 
In this situation, the reversal field of the recording bit does not act on the STO. 
Nevertheless, we believe that the above calculations are useful not only to the nanopillar structure 
but also to the commercial design of MAMR due to the following reason. 

In the commercially designed MAMR, the reversal field of the recording bit is generated from a writing pole \cite{zhu08}. 
The magnetic field generated from the writing pole has, however, a distribution around the recording head, 
and supplies a field acting on the STO in the recording head. 
In this situation, Eq. (\ref{eq:condition_Hc}) can still be used as a condition to generate microwave field from the STO, 
i.e., the stray field injected from the writing pole to the STO should be smaller than the critical value $H_{\rm c}$. 
We should again emphasize, however, that the reversal field of the recording bit in this situation is not restricted by $H_{\rm c}$ 
because the reversal field acting on the bit is different from the stray field acting on the STO. 





\section{Conclusion}
\label{sec:Conclusion}

The oscillation properties of an STO for MAMR is studied theoretically. 
It is shown that three dynamical phases, i.e., 
an in-plane magnetized state, out-of-plane self-oscillation state, and perpendicularly magnetized state, appear at zero temperature, 
depending on the current and applied field. 
The out-of-plane self-oscillation state applicable to MAMR 
disappears when the perpendicular field is larger than a critical value. 
The result indicates that the reversal field of a magnetic recording bit by MAMR should be smaller than the critical value 
when the type of STO investigated in this study is used as the microwave generator. 
The analytical formulas of currents characterizing the different dynamical phases are also obtained, 
which will be useful to estimate the value of the critical field theoretically. 
The analytical formulas indicate that a material with a small damping is not preferable 
to achieve a wide range of the self-oscillation applicable to MAMR, 
although such a material has been considered to be preferable from the view point of reduction of power consumption \cite{devolder13,tsunegi14}. 
The numerical simulation including stochastic noise indicates that 
the self-oscillation disappears even at finite temperature. 


\section*{Acknowledgements}

The authors express gratitude to Takehiko Yorozu, Satoshi Okamoto, Ryo Hiramatsu, Shingo Tamaru, and Ursula Ebels for valuable discussions. 
T.T. is thankful to Satoshi Iba, Aurelie Spiesser, Hiroki Maehara, and Ai Emura for their support and encouragement. 
This work is supported by the Japan Society and Technology Agency (JST) strategic innovation promotion program 
"Development of new technologies for 3D magnetic recording architecture".


\appendix

\section{Derivation of threshold current}
\label{sec:AppendixA}

In this Appendix, we show the derivations of the threshold current shown in the main text.


\subsection{Definition of threshold current}

The definition of the threshold current in this work is similar to that developed in our previous work \cite{taniguchi16}. 
In Ref. \cite{taniguchi16}, we studied the excitation of an out-of-plane self-oscillation in the same type of STO, 
where an in-plane anisotropy is zero and an external magnetic field points to an arbitrary direction. 
When the external magnetic field has a finite component in the in-plane direction, 
an energetically stable state appears around an in-plane axis. 
This stable state is separated from the out-of-plane state by the energy barrier between the stable state and a saddle point. 
In this case, a finite current is necessary to excite the out-of-plane self-oscillation by overcoming this energy barrier. 
In the present model, however, the external magnetic field does not have an in-plane component, 
but the in-plane anisotropy plays a similar role. 
The threshold current in the present study is defined as 
the minimum current necessary to move the magnetization from the stable state caused by the in-plane anisotropy to the out-of-plane state.



\begin{figure}
\centerline{\includegraphics[width=1.0\columnwidth]{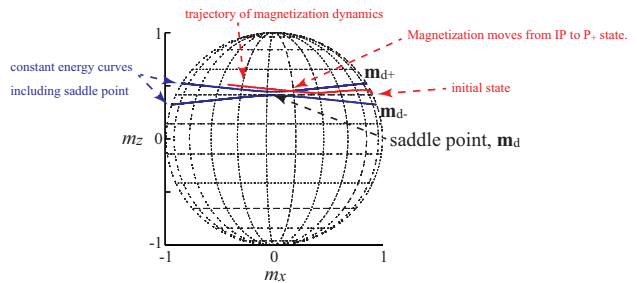}}
\caption{(Color online)
         A trajectory (red line) of the magnetization dynamics from an initial energetically stable state to an out-of-plane state, i.e., from the IP to P${}_{+}$ state, 
         where $H_{\rm appl}=7.0$ kOe and $j=-18.1$ MA/cm${}^{2}$. 
         The constant energy curves including the saddle point, separating the IP and P${}_{+}$ states, are also shown by the blue lines. 
         The saddle point is denoted as $\mathbf{m}_{\rm d}$, whereas points on the constant energy curve of the saddle point and satisfying $m_{y}=0$ are denoted as $\mathbf{m}_{{\rm d}\pm}$. 
         }
\label{fig:fig8}
\end{figure}



Figure \ref{fig:fig8} shows an example of the trajectory of the magnetization dynamics 
from the initial stable state to the out-of-plane state, i.e., from the IP to P${}_{+}$ state, by the red line. 
The value of the current density, $-18.1$ MA/cm${}^{2}$, is close to the threshold current found from the numerical simulation of the LLG equation. 
We also show constant energy curves including the saddle point, denoted as $\mathbf{m}_{\rm d}$, by the blue lines. 
As mentioned above, these lines separate the in-plane and out-of-plane states. 
As shown in Fig. \ref{fig:fig8}, the magnetization moves to the out-of-plane state without showing a precession around an in-plane direction. 
Note that the magnetization absorbs an energy from the work done by spin torque to overcome the energy barrier 
between the stable state and saddle point. 
We note that the magnetization crosses a point on the constant energy curve of $E_{\rm saddle}$ close to $\mathbf{m}_{\rm d}$, 
where $E_{\rm saddle}$ is the energy density of the saddle point. 
Thus, the energy change during the excitation is approximately described as 
\begin{equation}
  \int_{\mathbf{m}_{\rm stable}}^{\mathbf{m}_{\rm d}}
  dt 
  \frac{dE}{dt}
  \simeq
  E_{\rm saddle}
  -
  E_{\rm stable}, 
  \label{eq:energy_equation_orig}
\end{equation}
where $\mathbf{m}_{\rm stable}$ and $E_{\rm stable}$ are the energetically stable state and the energy density at this point. 
The exact solution of the LLG equation is necessary to solve this equation. 
However, the LLG equation is a nonlinear equation, and is difficult to solve exactly. 
Instead, the exact trajectory is approximated by a motion of the magnetization on a constant energy curve of $E_{\rm saddle}$, i.e., 
\begin{equation}
  \int_{\mathbf{m}_{{\rm d}\pm}}^{\mathbf{m}_{\rm d}}
  dt 
  \frac{dE}{dt}
  \simeq
  E_{\rm saddle}
  -
  E_{\rm stable}, 
  \label{eq:energy_equation}
\end{equation}
where $\mathbf{m}_{{\rm d}+}$ and $\mathbf{m}_{{\rm d}-}$ are the points on the constant energy curve of $E_{\rm saddle}$ and satisfying $m_{y}=0$; see Fig. \ref{fig:fig8}. 
The approximation from Eq. (\ref{eq:energy_equation_orig}) to (\ref{eq:energy_equation}) is valid 
when $\mathbf{m}_{\rm stable}$ locates close to $\mathbf{m}_{\rm d}$. 
This fact indicates that the in-plane anisotropy field $H_{\rm K}$ is smaller than the demagnetization field $4\pi M$ in the perpendicular direction.
As mentioned in Ref. \cite{taniguchi16}, 
the spin torque excited in the region of $[\mathbf{m}_{{\rm d}+},\mathbf{m}_{\rm d}]$ by a negative current 
has a component antiparallel to the damping torque. 
Therefore, the STO absorbs finite energy from the spin torque, and can overcome the damping torque. 
Similarly, the spin torque excited in the region of $[\mathbf{m}_{{\rm d}-},\mathbf{m}_{\rm d}]$ by a positive current 
has a component antiparallel to the damping torque, 
and thus, leads to an instability of the magnetization. 
Therefore, the current satisfying Eq. (\ref{eq:energy_equation}) is the threshold current density. 
Similar to Ref. \cite{taniguchi16}, the threshold current density is given by 
\begin{equation}
\begin{split}
  j_{{\rm th}\pm}
  =&
  \frac{2\alpha eMd}{\hbar \eta}
  \frac{\int_{\mathbf{m}_{{\rm d}\pm}}^{\mathbf{m}_{\rm d}} dt \left[\mathbf{H}^{2}-(\mathbf{m}\cdot\mathbf{H})^{2} \right]}
    {\int_{\mathbf{m}_{{\rm d}\pm}}^{\mathbf{m}_{\rm d}} dt [\mathbf{p}\cdot\mathbf{H}-(\mathbf{m}\cdot\mathbf{p})(\mathbf{m}\cdot\mathbf{H})]/(1+\lambda \mathbf{m}\cdot\mathbf{p})}
\\
  &+
  \frac{2ed}{\gamma \hbar \eta}
  \frac{E_{\rm saddle}-E_{\rm stable}}
    {\int_{\mathbf{m}_{{\rm d}\pm}}^{\mathbf{m}_{\rm d}} dt [\mathbf{p}\cdot\mathbf{H}-(\mathbf{m}\cdot\mathbf{p})(\mathbf{m}\cdot\mathbf{H})]/(1+\lambda \mathbf{m}\cdot\mathbf{p})}. 
  \label{eq:jth_def}
\end{split}
\end{equation}
According to the assumption mentioned above, 
the integrals are calculated along the lines on the constant energy curves of the saddle point energy. 
We introduce dimensionless quantities, 
\begin{equation}
  Q_{\alpha}
  =
  \frac{\gamma}{4\pi M}
  \int_{\mathbf{m}_{{\rm d}\pm}}^{\mathbf{m}_{\rm d}}
  dt 
  \left[
    \mathbf{H}^{2}
    -
    \left(
      \mathbf{m}
      \cdot
      \mathbf{H}
    \right)^{2}
  \right],
  \label{eq:Q_alpha}
\end{equation}
\begin{equation}
  Q_{\rm s}
  =
  \gamma
  \int_{\mathbf{m}_{{\rm d}\pm}}^{\mathbf{m}_{\rm d}}
  dt 
  \frac{\mathbf{p}\cdot\mathbf{H}-(\mathbf{m}\cdot\mathbf{p})(\mathbf{m}\cdot\mathbf{H})}{1+\lambda \mathbf{m}\cdot\mathbf{p}}. 
  \label{eq:Q_s}
\end{equation}
Before performing further calculations, let us analytically define the energetically stable state and the saddle point. 


\subsection{Minimum and saddle points of energy landscape}

For simplicity, we introduce following dimensionless quantities, 
\begin{align}
  \varepsilon
  =
  \frac{E}{4\pi M^{2}},
&&
  h
  =
  \frac{H_{\rm appl}}{4\pi M},
&&
  \kappa
  =
  \frac{H_{\rm K}}{4\pi M}.
\end{align}
We also introduce zenith and azimuth angles, $(\theta,\varphi)$, to represent the magnetization direction as 
$\mathbf{m}=(m_{x},m_{y},m_{z})=(\sin\theta\cos\varphi,\sin\theta\sin\varphi,\cos\theta)$. 
The dimensionless energy density $\varepsilon$ is expressed as 
\begin{equation}
  \varepsilon
  =
  -h 
  \cos\theta
  -
  \frac{\kappa}{2}
  \sin^{2}\theta
  \cos^{2}\varphi
  +
  \frac{1}{2}.
  \label{eq:energy}
\end{equation}

In the absence of an external field, the energetically stable states locate at $\mathbf{m}=\pm \mathbf{e}_{x}$. 
Even in the presence of a perpendicular field, the energetically stable states still locate at the positions satisfying $\varphi=0,\pi$. 
Consequently, the stable state is determined by a condition given by 
\begin{equation}
  \frac{\partial \varepsilon}{\partial \theta}
  \bigg|_{\varphi=0,\pi}
  =
  \left[
    h
    -
    \left(
      1
      +
      \kappa
    \right)
    \cos\theta
  \right]
  \sin\theta
  =
  0.
\end{equation}
When the magnitude of the applied field $h$ is smaller than $1+\kappa$, 
the stable state is 
\begin{equation}
  \theta_{\rm stable}
  =
  \cos^{-1}
  \frac{h}{1+\kappa},
\end{equation}
whereas $\theta=0,\pi$ correspond to the unstable states (energetically local maxima). 
On the other hand, when $h>1+\kappa$, $\theta=0$ becomes the stable state whereas $\theta=\pi$ is the unstable state. 
In the following, we assume that $h<1+\kappa$ because the threshold current $j_{{\rm th}\pm}$ is defined for this case. 
The stable states can be expressed as 
\begin{equation}
  \mathbf{m}_{\rm stable}
  =
  \begin{pmatrix}
    \pm \sqrt{1-[h/(1+\kappa)]^{2}} \\
    0 \\
    h/(1+\kappa)
  \end{pmatrix}.
\end{equation}
The energy at the stable state is given by 
\begin{equation}
  \varepsilon_{\rm stable}
  =
  -\frac{h^{2}+\kappa (1+\kappa)}{2(1+\kappa)}. 
\end{equation}

In contrast, saddle points locate at the points satisfying $m_{x}=0$. 
Then, the saddle points are determined from the condition 
\begin{equation}
  \frac{\partial \varepsilon}{\partial m_{z}}
  \bigg|_{m_{x}=0}
  =
  -h
  +
  m_{z}
  =
  0.
\end{equation}
Accordingly, the saddle points are given by 
\begin{equation}
  \mathbf{m}_{\rm d}
  =
  \begin{pmatrix}
    0 \\
    \pm \sqrt{1-h^{2}} \\
    h
  \end{pmatrix}.
\end{equation}
The energy at the saddle point is given by 
\begin{equation}
  \varepsilon_{\rm saddle}
  =
  -\frac{h^{2}}{2}.
\end{equation}
We find that the energy barrier between the stable state and the saddle point is given by 
\begin{equation}
  \Delta
  \varepsilon
  =
  \varepsilon_{\rm saddle}
  -
  \varepsilon_{\rm stable}
  =
  \frac{\kappa}{2}
  \left(
    1
    -
    \frac{h^{2}}{1+\kappa}
  \right). 
\end{equation}

For the sake of calculations which appear later in the discussion, 
we introduce a notation 
\begin{equation}
  z_{\rm d}
  \equiv 
  \mathbf{m}_{\rm d}
  \cdot
  \mathbf{e}_{z}
  =
  h.
\end{equation}
Also, we use the following notation, 
\begin{equation}
  z_{{\rm d}\pm}
  \equiv 
  \mathbf{m}_{{\rm d}\pm}
  \cdot
  \mathbf{e}_{z}
  =
  \frac{h \pm \sqrt{\kappa (1-h^{2}+\kappa)}}{1+\kappa}.
\end{equation}
Here, $z_{{\rm d}\pm}$ is obtained from a condition 
$\varepsilon(m_{y}=0)=\varepsilon_{\rm saddle}$. 


\subsection{Derivation of threshold current}

Now let us return to the calculation of the threshold current. 
Using the dimensionless quantities, Eq. (\ref{eq:jth_def}) can be rewritten as 
\begin{equation}
  j_{{\rm th}\pm}
  =
  \frac{2e Md}{\hbar \eta}
  4\pi M 
  \frac{\Delta \varepsilon + \alpha Q_{\alpha}}{Q_{\rm s}}.
  \label{eq:jth_Appendix}
\end{equation}

We note that a constant energy curve corresponds to an orbit determined from the Landau-Lifshitz equation, $d \mathbf{m}/dt=-\gamma \mathbf{m}\times \mathbf{H}$. 
From this equation, we transform the integral variable in Eqs. (\ref{eq:Q_alpha}) and (\ref{eq:Q_s}) from $t$ to $m_{z}$ 
through the relation 
\begin{equation}
  dt 
  =
  \frac{dm_{z}}{\gamma H_{\rm K} m_{x} m_{y}}.
\end{equation}
Note that $m_{x}$ can be expressed in terms of $\varepsilon$ and $m_{z}$ as 
\begin{equation}
  m_{x}
  =
  \pm
  \sqrt{
    \frac{-2 \varepsilon + m_{z}(m_{z}-2h)}{\kappa}
  }, 
\end{equation}
where we use Eq. (\ref{eq:energy}). 
In particular, on the line including the saddle point, 
\begin{equation}
  m_{x}(\varepsilon=\varepsilon_{\rm saddle})
  =
  \pm
  \sqrt{
    \frac{(h-m_{z})^{2}}{\kappa}
  }. 
  \label{eq:mx}
\end{equation}
Then, $m_{y}$ can be expressed in terms of $m_{z}$ as 
\begin{equation}
\begin{split}
  m_{y}
  &=
  \pm
  \sqrt{
    1
    -
    m_{x}^{2}
    -
    m_{y}^{2}
  }
\\
  &=
  \pm
  \sqrt{
    \frac{-(1+\kappa) m_{z}^{2} + 2 h m_{z} + \kappa - h^{2}}{\kappa}
  }. 
  \label{eq:my}
\end{split}
\end{equation}

Using these $m_{x}$ and $m_{y}$, we notice that $Q_{\alpha \pm}$ and $Q_{{\rm s}\pm}$ can be expressed as 
\begin{equation}
  Q_{\alpha \pm}
  =
  \int_{z_{{\rm d}\pm}}^{z_{\rm d}}
  dm_{z}
  \frac{\pm (1-h^{2}+\kappa)(h-m_{z})^{2}}{\sqrt{(h-m_{z})^{2}[-(1+\kappa)m_{z}^{2}+2hm_{z}+\kappa-h^{2}]}},
\end{equation}
\begin{equation}
\begin{split}
  &
  Q_{{\rm s}\pm}
  =
\\
  &
  \int_{z_{{\rm d}\pm}}^{z_{\rm d}}
  dm_{z}
  \frac{\mp (h-m_{z}) (1-hm_{z})}{(1+\lambda m_{z}) \sqrt{(h-m_{z})^{2}[-(1+\kappa)m_{z}^{2}+2hm_{z}+\kappa-h^{2}]}}.
\end{split}
\end{equation}
Note that $m_{z}$ in the calculations of $Q_{\alpha+}$ and $Q_{{\rm s}+}$ is in the range of $z_{\rm d}=h \le m_{z} \le z_{{\rm d}+}$, 
whereas that for $Q_{\alpha-}$ and $Q_{{\rm s}-}$ is in the range of $z_{{\rm d}-} \le m_{z} \le z_{\rm d}=h$. 
Therefore, a factor $\sqrt{(h-m_{z})^{2}}$ in the denominators of $Q_{\alpha\pm}$ and $Q_{{\rm s}\pm}$ should be 
$m_{z}-h$ for the calculations of $Q_{\alpha+}$ and $Q_{{\rm s}+}$, 
and $h-m_{z}$ for the calculations of $Q_{\alpha-}$ and $Q_{{\rm s}-}$. 

We note that the following integral formulas are useful to calculate $Q_{\alpha\pm}$ and $Q_{{\rm s}\pm}$; 
\begin{equation}
\begin{split}
  I_{{\rm F}0}
  &\equiv
  \int
  \frac{dx}{\sqrt{ax^{2}+bx+c}}
\\
  &=
  -\frac{1}{\sqrt{-a}}
  \sin^{-1}
  \left(
    \frac{2ax+b}{\sqrt{b^{2}-4ac}}
  \right),
\end{split}
\end{equation}
\begin{equation}
\begin{split}
  I_{{\rm F}1}
  &\equiv
  \int dx 
  \frac{x}{\sqrt{ax^{2}+bx+c}}
\\
  &=
  \frac{\sqrt{ax^{2}+bx+c}}{a}
  -
  \frac{b}{2a}
  I_{{\rm F}0}, 
\end{split}
\end{equation}
\begin{equation}
\begin{split}
  I_{{\rm F}2}
  &\equiv
  \int 
  \frac{dx}{(px+q) \sqrt{ax^{2}+bx+c}}
\\
 &
  =
  \frac{1}{\sqrt{|k|}}
  \sin^{-1}
  \left[
    \frac{(bp-2qa)(px+q) + 2k}{p|px+q|\sqrt{b^{2}-4ac}}
  \right],
\end{split}
\end{equation}
where $x=m_{z}$, $k=aq^{2}-bpq+cp^{2}$, $a=-(1+\kappa)$, $b=2h$, $c=\kappa-h^{2}$, $q=1$, and $p=\lambda$ for the present calculations. 
The formula of $I_{{\rm F}0}$ is valid when $a<0$, 
whereas that of $I_{{\rm F}2}$ is valid when $k<0$
We also note that 
\begin{equation}
  \frac{1-h m_{z}}{1+\lambda m_{z}}
  =
  -\frac{h}{\lambda}
  +
  \frac{1+(h/\lambda)}{1+\lambda m_{z}}. 
\end{equation}
Using these relations, we find that 
\begin{equation}
\begin{split}
  Q_{\alpha\pm}
  &=
  \int_{z_{{\rm d}\pm}}^{z_{\rm d}}
  dm_{z}
  \frac{(1-h^{2}+\kappa)(h-m_{z})}{\sqrt{-(1+\kappa)m_{z}^{2}+2hm_{z}+\kappa-h^{2}}}
\\
  &=
  \left(
    1
    -
    h^{2}
    +
    \kappa
  \right)
  \left\{
    \frac{\sqrt{\kappa(1-h^{2})}}{1+\kappa}
  \right.
\\
  &
  \left.
  \ \ \ \ \ \ 
    \mp
    \frac{\kappa h}{(1+\kappa)^{3/2}}
    \cos^{-1}
    \left[
      \frac{\pm \kappa h}{\sqrt{\kappa(1-h^{2}+\kappa)}}
    \right]
  \right\},
  \label{eq:Q_alpha_sol}
\end{split}
\end{equation}
\begin{equation}
\begin{split}
  Q_{{\rm s}\pm}
  &=
  \int_{z_{{\rm d}\pm}}^{z_{\rm d}}
  d m_{z}
  \frac{(1-hm_{z})}{(1+\lambda m_{z})\sqrt{-(1+\kappa)m_{z}^{2}+2hm_{z}+\kappa -h^{2}}}
\\
  &=
  \int_{z_{{\rm d}\pm}}^{z_{\rm d}}
  d m_{z}
  \frac{-(h/\lambda) + (1+h/\lambda)(1+\lambda m_{z})}{\sqrt{-(1+\kappa)m_{z}^{2}+2hm_{z}+\kappa -h^{2}}}
\\
  &=
  \pm
  \frac{h \cos^{-1} \left[ \frac{\pm \kappa h}{\sqrt{\kappa(1-h^{2}+\kappa)}} \right]}{\lambda \sqrt{1+\kappa}}
\\
  &\ \ \ \ 
  \mp
  \frac{(h+\lambda) \cos^{-1} \left[ \frac{\pm \kappa(h+\lambda)}{(1+\lambda h) \sqrt{\kappa(1-h^{2}+\kappa)}} \right]}{\lambda \sqrt{(1+\lambda h)^{2}+\kappa(1-\lambda^{2})}}.
  \label{eq:Q_s_sol}
\end{split}
\end{equation}
Substituting Eqs. (\ref{eq:Q_alpha_sol}) and (\ref{eq:Q_s_sol}) into Eq. (\ref{eq:jth_Appendix}), 
the theoretical formula of the threshold current is obtained. 
We note that $I_{\alpha\pm}$ and $I_{{\rm s}\pm}$ in Eq. (\ref{eq:jth}) correspond to 
$Q_{\alpha\pm}/(1-h^{2}+\kappa)$ and $Q_{{\rm s}\pm}$ in this Appendix, respectively. 


\subsection{Special limits}

In the case of $h=0$, the threshold current becomes 
\begin{equation}
\begin{split}
  j_{{\rm th}\pm}(h=0)
  &=
  \mp
  \frac{2eMd}{\hbar \eta}
  4\pi M
  \frac{\sqrt{\kappa[1+(1-\lambda^{2})\kappa]} (\sqrt{\kappa} + 2 \alpha)}{2 \cos^{-1}[\pm \lambda \sqrt{\kappa/(1+\kappa)}]}
\\
  &\to
  \mp
  \frac{2eMd}{\pi \hbar \eta}
  4\pi M 
  \sqrt{\kappa (1+\kappa)} 
  \left(
    \sqrt{\kappa} 
    + 
    2\alpha
  \right),
  \label{eq:jth_limit}
\end{split}
\end{equation}
where the limit of $\lambda \to 0$ is considered at the last line. 

We should mention here that another approach evaluating the instability threshold of an in-plane magnetized free layer was developed in Ref. \cite{bazaliy12}, 
where the LLG equation is simplified by assuming that the cone angle of the magnetization is close to 90${}^{\circ}$. 
The instability threshold was studied in terms of the effective energy and damping. 
Equation (30) of Ref. \cite{bazaliy12} with $\theta_{s}=0$ looks similar to the second line of Eq. (\ref{eq:jth_limit}) of the present work, 
except for a minor difference of the numerical factor in the denominator, which is $2$ in Ref. \cite{bazaliy12} and $\pi$ in Eq. (\ref{eq:jth_limit}), 
where $\theta_{s}$ is the zenith angle of the pinned layer. 
A difference between our theory and that of Ref. \cite{bazaliy12} is that the present work is applicable to an STO in the presence of a perpendicular field, as shown in Fig. \ref{fig:fig4}, 
whereas the situation studied in Ref. \cite{bazaliy12} can be used to the case of $h=0$ only due to the assumption that the zenith angle is close to 90 degree. 
It might be possible to extend the theory of Ref. \cite{bazaliy12} to an STO with the perpendicular field 
by expanding the LLG equation around a different zenith angle. 
Investigating such a possibility is, however, not the main subject of this paper. 


\subsection{The other current scales}

In the main text, the other current scales, $j_{\rm c}$ and $j_{{\rm u}\pm}$ appear. 
Here, $j_{\rm c}$ is the critical current density destabilizing the magnetization in equilibrium. 
On the other hand, $j_{{\rm u}\pm}$ is a current necessary to move the magnetization parallel to the $z$ axis, i.e., $\mathbf{m}=\pm\mathbf{e}_{z}$. 


These currents are obtained from the linearized LLG equation. 
In the linearized LLG equation, we assume that the magnetization $\mathbf{m}$ is nearly parallel to an axis $\mathbf{n}$, 
and shows a small amplitude oscillation around this axis. 
Then, we introduce a transverse magnetization $\delta\mathbf{m}$ as $\mathbf{m} \simeq \mathbf{n}+\delta \mathbf{m}$, 
and LLG equation is expressed in the first-order terms of $\delta\mathbf{m}$. 
The solution of $\delta \mathbf{m}$ can be expressed as $\delta\mathbf{m} \propto e^{-i \omega t}$. 
When ${\rm Im}[\omega]<0$, $\delta \mathbf{m}$ decays to zero, 
whereas when ${\rm Im}[\omega]>0$, $\delta \mathbf{m}$ increases with increasing time. 
Therefore, a condition ${\rm Im}[\omega]=0$ determines the stability of the magnetization. 
We should note that the linearized LLG equation is applicable only to an oscillation with a small amplitude, i.e., $|\delta\mathbf{m}| \ll 1$. 
Thus, in general, the linearized LLG equation cannot be applied to a self-oscillation state, 
such as an out-of-plane self-oscillation shown in Fig. \ref{fig:fig2}(b) of the main text. 
However, the LLG equation can be applied to investigate the stability of an energetically stable state in the present system 
because this state is now just a point, and thus, an oscillation around the stable state can be regarded as a small-amplitude oscillation. 
We also notice that the linearized LLG equation is applicable to an energetically unstable states $\mathbf{m}=\pm\mathbf{e}_{z}$. 


The details of the derivation of the critical current density for a free layer 
having an arbitrary uniaxial anisotropy and external field are summarized in Appendix A of our previous work \cite{taniguchi16}. 
As mentioned above, the energetically stable states locate at $\varphi=0,\pi$. 
On the other hand, $\theta$ corresponding to the stable state is $\theta=\cos^{-1}[h/(1+\kappa)]$, as mentioned above. 
Substituting these relations to equations in Appendix A of Ref. \cite{taniguchi16}, the critical current density is given by Eq. (\ref{eq:jc}). 

We also apply the linearized LLG equation to calculate $j_{{\rm u}\pm}$. 
In this case, $\theta$ in the equations of Ref. \cite{taniguchi16} is zero for $j_{{\rm u}+}$ and $\pi$ for $j_{{\rm u}-}$. 
Then, we find that $j_{{\rm u}\pm}$ is given by Eq. (\ref{eq:ju}). 


In the present system, the spin torque is finite at the energetically stable point, $\mathbf{m}_{\rm stable}$. 
Therefore, one might consider that the linearization of the LLG equation should be done around the steady point determined by the sum of all torques, 
not around the stable state, to evaluate $j_{\rm c}$. 
In this case, the zenith and azimuth angles $(\theta_{0},\varphi_{0})$ of the static point, 
which are necessary to evaluate the critical current, depend on the current magnitude. 
Therefore, two equations determining the steady point and the critical current formula should be solved self-consistently with respect to $(\theta_{0},\varphi_{0})$ and $j_{\rm c}$; 
see, for example, Eqs. (6), (7), and (A7) in Ref. \cite{taniguchi13JAP}. 
In this case, it is difficult to obtain a simple analytical formula of $j_{\rm c}$. 
We notice that, at least for STO studied in this paper, the evaluation of $j_{\rm c}$ using $\mathbf{m}_{\rm stable}$, instead of the steady point, 
is a good approximation. 
For example, in Ref. \cite{taniguchi16}, the critical current, $j_{\rm c}$, for a tilted magnetic field without the in-plane anisotropy was estimated to be $328 \times 10^{6}$ A/cm${}^{2}$, 
where the stable point was used for the evaluation. 
The value changes to $280 \times 10^{6}$ A/cm${}^{2}$ when the stable state is replaced by the steady point in the evaluation of $j_{\rm c}$. 
We emphasize that both of them are two orders of magnitude larger than the instability threshold found in the numerical simulation, $7.3 \times 10^{6}$ A/cm${}^{2}$. 
Therefore, the linearization of the LLG equation does not work to evaluate the instability threshold even after replacing the stable state used in the evaluation of $j_{\rm c}$ to the static point. 
We should remind the readers that this is because the initial state that should be adopted in our simulation is the energetically stable state and not the static point, 
which is also reasonably in accordance with the conditions achieved in experiments; see Sect. \ref{sec:Applicability of present formula}. 


We also note that there is another approach to define the instability threshold \cite{sodemann11}. 
Here, the instability of the steady point occurs when the steady point collides with the saddle point, and both of them disappear. 
The threshold current estimated by this method is denoted as $J_{{\rm c}1}$ in Ref. \cite{ebels08}. 
We note that such current $J_{{\rm c}1}$ is independent of the damping constant $\alpha$ 
because it is determined from the steady state solution of the LLG equation, 
i.e., with the term $d\mathbf{m}/dt$ being set to zero, the damping constant $\alpha$ does not appear in the equation determining $J_{{\rm c}1}$. 
The role of $J_{{\rm c}1}$ on the instability threshold is discussed in Ref. \cite{ebels08}. 
Note that $J_{{\rm c}1}$ does not appear in the present study. 
This is because the initial state in our study is far away from the steady point, as explained in Sect. \ref{sec:Applicability of present formula}, 
and therefore, the magnetization can move to the self-oscillation state by avoiding this point. 
It should also be reminded that the theoretical analysis of the instability threshold based on $J_{{\rm c}1}$ has been used in different systems. 
In Refs. \cite{liu12,lee13}, for example, 
the switching of a perpendicularly magnetized ferromagnet by the spin Hall effect is studied, 
where the threshold current is determined from the condition that the steady state solution of the LLG equation disappears. 
A comparison of the instability threshold based on $J_{{\rm c}1}$ \cite{liu12,lee13} and $j_{\rm c}$ for the spin Hall geometry is discussed in Ref. \cite{taniguchi15PRB}.


\subsection{For a large $\lambda$ case}
\label{sec:For a large lambda case}

Here, we briefly discuss the role of the parameter $\lambda$ characterizing the spin torque asymmetry. 
The parameter $\lambda$ originates from the spin-dependent transport in ferromagnetic multilayers. 
For example, in a metallic system, the amount of the spin accumulation near the antiparallel alignment of the magnetization is 
larger than that near the parallel alignment. 
Therefore, the spin torque near the antiparallel alignment of $\mathbf{m}$ and $\mathbf{p}$ becomes stronger than that near the parallel alignment \cite{slonczewski96,slonczewski05,xiao04,xiao07}. 
The parameter $\lambda$ is an index to describe such asymmetry, and is in the range of $0 \le \lambda < 1$. 
For example, in a symmetric magnetic tunnel junction, $\lambda$ is a square of the spin polarization \cite{slonczewski05}, 
and therefore, is usually small. 
In fact, the parameter $\lambda$ has often been neglected in the analysis of the magnetization dynamics excited by spin torque, 
although it has been shown that this parameter plays a key role on the magnetization dynamics even if it is small \cite{taniguchi13}. 
The present study also assumes a small $\lambda=\eta^{2}=0.36$. 
For the sake of generality, however, let us consider the role of $\lambda$ discussed in the previous sections. 

The self-oscillation which is discussed here can be approximated as depicting 
a trajectory on a constant energy curve of $E=-M \int d \mathbf{m}\cdot\mathbf{H}$ 
because the self-oscillation is stabilized when the dissipation due to the damping torque balances the energy supplied by the spin torque. 
Therefore, the current necessary to excite a self-oscillation on a certain trajectory is identified by energy density $E$. 
Let us denote such current density as $j(E)$. 
In the present system where $\kappa$ is finite, 
a very complex procedure is necessary to derive an analytical formula of $j(E)$ for an arbitrary $E$. 
It should however be noted that $j_{{\rm u}\pm}$ in Eq. (\ref{eq:ju}) can be defined as \cite{taniguchi16} 
\begin{equation}
  j_{{\rm u}\pm}
  =
  \lim_{E \to E_{{\rm max}\pm}}
  j(E), 
\end{equation}
where $E_{{\rm max}\pm}$ is the local maximum of $E$ at $\mathbf{m}=\pm\mathbf{e}_{z}$. 

In general case, the out-of-plane self-oscillation is excited when the threshold current $j_{{\rm th}\pm}$ satisfies the condition $j_{{\rm th}\pm}/{\rm max}[j(E)]<1$. 
For a small $\lambda$, $j(E)$ is expected to be a monotonically increasing function of $E$. 
Under this condition, we can replace ${\rm max}[j(E)]$ with $j_{{\rm u}\pm}$. 
Therefore, OP${}_{\pm}$ states appear when $j_{{\rm th}\pm}/j_{{\rm u}\pm}<1$, as mentioned in Sect. \ref{sec:Theoretical formulas of characteristic currents}.
In fact, as shown in the following sections, the condition $j_{{\rm th}\pm}/j_{{\rm u}\pm}<1$ works well to identify the OP state in the present system, 
indicating that $j(E)$ is a monotonic function of $E$. 

The conclusion may change for a large $\lambda$. 
To make this point clear, let us consider a free layer 
where both the in-plane anisotropy field $H_{\rm K}$ and the applied field $H_{\rm appl}$ are zero, for simplicity. 
In this case, the free layer becomes highly symmetric with respect to the $z$ axis, 
and the energy density $E=2\pi M^{2}\cos^{2}\theta$ is identified by the cone angle $\theta=\cos^{-1}m_{z}$ from the $z$ axis. 
Then, the explicit form of $j(E)$ is obtained as 
\begin{equation}
  j(\theta)
  =
  -\frac{2 \alpha (1+\lambda \cos\theta) eMd}{\hbar \eta}
  4\pi M 
  \cos\theta.
  \label{eq:j_theta}
\end{equation}
We note that $j(\theta)$ at the energetically maxima states locating at $\theta=0,\pi$ become 
\begin{equation}
  j(\theta\to 0,\pi)
  =
  \mp
  \frac{2 \alpha (1 \pm \lambda) eMd}{\hbar \eta}
  4\pi M,
\end{equation}
which correspond to Eq. (\ref{eq:ju}) in the limit of $\kappa,h \to 0$. 
We note that when $\lambda<1/2$, $|j(\theta)|$ monotonically increases where $\theta$ varies from $\pi/2$ to $0$ or $\pi$. 
In this case, $j_{{\rm u}+}$ ($j_{{\rm u}-}$) is given by $j(\theta=0)$ [$j(\theta=\pi$)]. 
This means that ${\rm max}[j(E)]=j_{{\rm u}\pm}$ mentioned above is satisfied. 
On the other hand, when $\lambda$ is larger than $1/2$, 
$j(\theta)$ for $\pi/2 \le \theta \le \pi$ has a maximum at an angle $\theta^{*}=\cos^{-1}[-1/(2\lambda)]$, 
whereas that for $0 \le \theta \le \pi/2$ is maximized at $\theta=0$. 
In other words, ${\rm max}[j(\theta)]$ cannot be explicitly given by $j_{{\rm u}-}$. 

Although this provides an example for a simplified case, i.e., both $\kappa$ and $h$ are zero, 
a similar situation may appear in the present system, 
i.e., $j(E)$ for a large $\lambda$ may have a maximum at a certain point $E^{*} \neq E_{{\rm max}\pm}$. 
In such case, the condition to excite the self-oscillation should be replaced 
from $j_{{\rm th}\pm}/j_{{\rm u}\pm}<1$ to $j_{{\rm th}\pm}/{\rm max}[j(E)]<1$. 
A further quantification of this condition requires the calculation of $j(E)$ for the present system, which is beyond the scope of this paper. 
At the same time, we should again emphasize that ${\rm max}[j(E)]$ in the present parameters is $j_{{\rm u}\pm}$, 
and therefore, the conclusion in Sect. \ref{sec:Theoretical formulas of characteristic currents} is unchanged in the main text.


\section{Another example on the dependence of instability threshold on initial state}
\label{sec:AppendixB}

As mentioned in Sect. \ref{sec:Applicability of present formula}, the instability threshold of the STO depends on the initial state. 
A similar example was found in the problem of the instability threshold in MAMR \cite{taniguchi14}. 
The theoretical analysis of the linearized LLG equation of MAMR in the presence of microwave field was developed in Ref. \cite{bertotti01}, 
and showed good agreement with the numerical simulation in Ref. \cite{okamoto10}. 
In Ref. \cite{okamoto10}, the microwave field is applied to a perpendicularly magnetized recording bit before applying a direct field \cite{comment_okamoto}. 
Therefore, the magnetization first moves to a steady point, which is slightly tilted from the perpendicular axis due to the in-plane microwave field. 
Then, the direct field is applied in reversed direction. 
When the field magnitude becomes larger than a threshold value, the magnetization reversal occurs. 
This case is similar to the instability threshold of the STO studied in Ref. \cite{ebels08}. 

On the other hand, a different situation can be expected both in experiments and simulation \cite{zhu08}. 
That is the case where both reversed field and microwave are applied simultaneously to recording bit. 
In this case, the magnetization reverses its direction before moving to a steady point. 
The instability threshold for such example was investigated in Ref. \cite{taniguchi14}, 
where a good agreement between the theory and numerical simulation was obtained. 

As emphasized in Ref. \cite{taniguchi14},  the theoretical studies made in both situations are useful to study the instability threshold of MAMR. 
The difference between them is related to the initial state of the magnetization, 
which is dependent on how the direct field and the microwave are applied. 


\section{Analytical formula of $H_{\rm c}$ in small damping limit}
\label{sec:AppendixC}

The critical field $H_{\rm c}$ in the small damping limit is obtained by solving Eq. (\ref{eq:condition_Hc}) with respect to the applied field $h=H_{\rm appl}/(4\pi M)$. 
It is, however, hardly possible to obtain an exact solution of $H_{\rm c}$ from Eq. (\ref{eq:condition_Hc}). 
Therefore, let us investigate an approximated formula of $H_{\rm c}$. 

As mentioned in Sect. \ref{sec:Conditions to observe MAMR}, 
the in-plane anisotropy field $H_{\rm K}$ should be small as much as possible to achieve a wide range of the self-oscillation state for MAMR. 
In fact, for example, $\kappa=H_{\rm K}/(4\pi M) \simeq 0.01$ is small. 
Moreover, $j_{{\rm th}+}$ in Eq. (\ref{eq:jth}) is dominated by a damping-independent term in the small-damping limit. 
Then, up to the first order of $\kappa=H_{\rm K}/(4\pi M)$ and in the limit of $\alpha \to 0$, 
the threshold current density $j_{{\rm th}+}$ becomes 
\begin{equation}
  j_{{\rm th}+}
  \simeq
  -\frac{2 e Md}{\pi \hbar \eta}
  4\pi M 
  \left(
    1
    +
    \lambda h
  \right) 
  \kappa.
\end{equation}
Solving Eq. (\ref{eq:condition_Hc}) with this approximated $j_{{\rm th}+}$, 
the approximated formula of the critical field is 
\begin{equation}
  H_{\rm c}
  \simeq
  \frac{[\pi \alpha (1+\lambda)(2+\kappa) - 2 \kappa]}{2[\lambda \kappa + \pi \alpha (1+\lambda)]}
  4\pi M. 
  \label{eq:Hc_approx}
\end{equation}
In the limit of $\kappa \to 0$, $H_{\rm c}$ becomes $4 \pi M$, 
indicating that the out-of-plane self-oscillation (OP${}_{+}$) state always appears for the field $H_{\rm appl}$ smaller than the demagnetization field, 
as mentioned in Sect. \ref{sec:Conditions to observe MAMR}. 
The value of Eq. (\ref{eq:Hc_approx}) for $\alpha=0.01$ is 10.66 kOe, 
which is larger than the exact value, 9.23 kOe. 




\end{document}